\documentclass[9pt]{extarticle}
\usepackage{spconf,amsmath,graphicx}

\usepackage{enumitem}
\setlist{nosep, leftmargin=14pt}
\usepackage{graphicx}
\usepackage{cite}
\usepackage{acronym}
\usepackage{multirow}
\usepackage{algorithm2e}
\usepackage{amsmath,amssymb,amsfonts}
\usepackage{subcaption}
\usepackage{cancel}
\usepackage{multirow}
\usepackage[normalem]{ulem}
\usepackage{tabularx}
\usepackage[hidelinks]{hyperref}
\usepackage{url}
\usepackage{tikz}
\usepackage[strings]{underscore}
\usepackage{xr}
\useunder{\uline}{\ul}{}
\newcolumntype{Y}{>{\centering\arraybackslash}X}

\DeclareMathOperator*{\argmin}{arg\,min}
\title{Single Image Compressed Sensing MRI via a Self-Supervised Deep Denoising Approach}
\name{Marlon Bran Lorenzana, Feng Liu and Shekhar S. Chandra}
\address{The University of Queensland\\
School of Electrical Engineering and Computer Science\\
St. Lucia}
\begin{document}

\maketitle              % typeset the header of the contribution
\begin{abstract}

Popular methods in compressed sensing (CS) are dependent on deep learning (DL), where large amounts of data are used to train non-linear reconstruction models. However, ensuring generalisability over and access to multiple datasets is challenging to realise for real-world applications. To address these concerns, this paper proposes a single image, self-supervised (SS) CS-MRI framework that enables a joint deep and sparse regularisation of CS artefacts. The approach effectively dampens structured CS artefacts, which can be difficult to remove assuming sparse reconstruction, or relying solely on the inductive biases of CNN to produce noise-free images. Image quality is thereby improved compared to either approach alone. Metrics are evaluated using Cartesian 1D masks on a brain and knee dataset, with PSNR improving by 2-4dB on average.  

% Popular methods in compressed sensing (CS) are dependent on deep learning (DL), where large amounts of data are used to train non-linear reconstruction models. However, the difficulty to generalise over and acquire multiple datasets make these methods challenging to realise for real-world applications. To address these concerns, this paper proposes a single image, self-supervised (SS) CS-MRI framework that can outperform and synergise with popular hand-crafted CS and single image SS-DL algorithms. The approach enables a joint deep and sparse regularisation of CS artefacts, without pre-training or ground truth images. We find the combined regularisation effectively dampens structured CS artefacts, which can be difficult to remove assuming a sparse reconstruction model. Performance is evaluated using Cartesian 1D masks on a brain and knee dataset, with PSNR improving by 2-4dB on average.

\end{abstract}


\begin{figure*}[t]
    \centering
    \includegraphics[width=\textwidth]{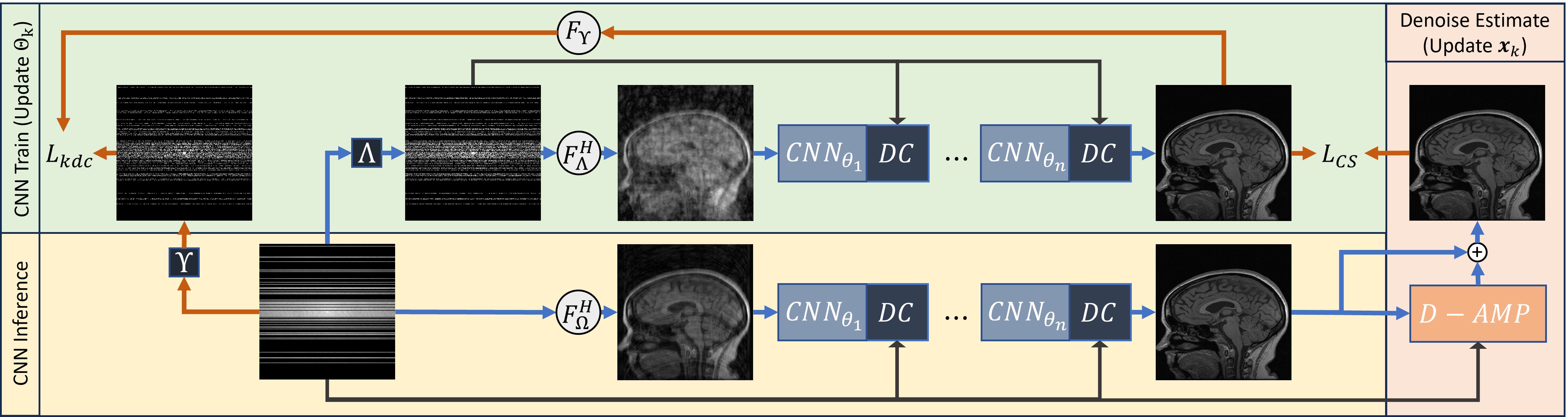}
    \caption{Proposed single image, self-supervised training procedure. Each epoch, sampled $k$-space $\mathbf{y}_\Omega$ is divided into subsets $\mathbf{y}_\Lambda, \mathbf{y}_\Upsilon$. Image estimates $\hat{\mathbf{x}}_\Lambda = f(F_\Lambda^H\mathbf{y}_\Lambda | \Theta_k)$ and $\hat{\mathbf{x}}_\Omega = f(F_\Omega^H\mathbf{y}_\Omega|\Theta_k)$ are then generated. \acs{CNN} and \acs{DC} blocks are described in Eq.~\ref{eq:dccnn}, where $\Theta_k = \{\theta_t\}_{t=1}^n$ at epoch $k$. The \acs{MSE} between $\hat{\mathbf{x}}_\Lambda$ and the denoiser estimate $\mathbf{x}_k$ ($L_{CS} := \frac{\mu}{2} \|\mathbf{x}_k - \hat{\mathbf{x}}_\Lambda - \mathbf{q}_k \|_2^2$), as well as withheld $k$-space $\mathbf{y}_\Upsilon$ ($L_{kdc} := \| \mathbf{y}_\Upsilon - F_\Upsilon \mathbf{\hat{x}}_\Lambda \|_2^2$), are used to update $\mathrm{\Theta}$ (Eq.~\ref{eq:cnn}). $\mathbf{q}_k$ is the Lagrange multipliers vector. Rather than relying on the inherent bias of \acs{CNN} to produce noise-free images via $L_{kdc}$, inclusion of $L_{CS}$ ensures the solution is noise-free with respect to the chosen \acs{CS}-\acs{MRI} algorithm.}
    \label{fig:self-supervise}
\end{figure*}

\section{Introduction}
\Ac{MRI} is a medical imaging technique that captures detailed cross-sectional images of the body without ionizing radiation. However, the process of acquiring data is slow compared to other modalities due to physical and biological factors. A popular method of acceleration is via $k$-space (discrete Fourier space) under-sampling. Importantly, sampling schemes are constrained to continuous trajectories through $k$-space, which can lead to structured image artefacts. Many hand-crafted recovery techniques have been proposed that produce high-quality images, relevant for this work are algorithms based on \ac{CS}~\cite{lustig_sparse_2007, metzler_denoising_2016, ye_compressed_2019}. Recently, the advent of \ac{DL} has delivered significant contributions for medical image recovery, enabling a data-driven approach to solve difficult inverse problems~\cite{schlemper_deep_2017, chandra_deep_2021, lorenzana_aliasnet_2023}. 

%To that end, recent approaches deploy a \ac{PnP} methodology that incorporates a \ac{DL} denoiser into an iterative recovery procedure~\cite{ahmad_plug-and-play_2020}. Though \ac{DL}-\ac{PnP} outperforms hand-crafted alternatives, the denoising network typically requires pre-training from ground-truth images.
Although image quality has benefited from \ac{DL}, there are still questions regarding generalisability over multiple datasets, considering fully-sampled training data is not always guaranteed. \Ac{SS} procedures mitigate this deficiency by training \ac{DL} \ac{CS}-\ac{MRI} models from large datasets of under-sampled measurements~\cite{yaman_self-supervised_2020, zeng_review_2021, hu_self-supervised_2021, zhou_dual-domain_2022}. However, out of domain reconstruction remains a concern~\cite{yaman_zero-shot_2022}. Foregoing training examples altogether, \ac{DIP} exploits the inductive biases of \ac{CNN} architectures to solve single image inverse problems, leveraging the relationship between the reconstructed image and the image degradation model~\cite{ulyanov_deep_2018}. ConvDecoder~\cite{darestani_accelerated_2021} and \ac{ZS-SSL}~\cite{yaman_zero-shot_2022} are examples of \ac{DIP}-style solutions to \ac{CS}-\ac{MRI}. Importantly, \ac{ZS-SSL} improves upon \ac{DIP} via a loss function that is reminiscent of Self2Self With Dropout~\cite{quan_self2self_2020} and variational recovery; convergence guarantees remain a topic of interest~\cite{yaman_zero-shot_2022}. The robustness of \ac{DIP} was later improved with DeepRED~\cite{mataev_deepred_2019}, which combines with \ac{RED}~\cite{romano_little_2017} to help guide the reconstruction process. While DeepRED is capable of producing high quality images, performance is dependent on the suitability of the chosen denoiser to the image degradation model. Unfortunately, structured \ac{CS} artefacts are often difficult to remove assuming a Gaussian noise distribution~\cite{metzler_denoising_2016}.

Recognising that sparse regularisation is effective for \ac{CS}~\cite{ye_compressed_2019}, this work proposes a deep and sparse \ac{SS}-\ac{CS}-\ac{MRI} framework that improves the regularisation of single image \ac{SS} \ac{CNN} without requiring training data. Our algorithm can be described as a DeepRED-style approach, where a \ac{CNN} prior is trained on the degraded measurements (similar to \ac{DIP}), while a hand-crafted \ac{CS} algorithm guides the reconstruction. Inclusion of the \ac{CS} algorithm ensures robustness in the final solution, as hand-crafted techniques are characterised by recovery guarantees and predictable behaviour~\cite{donoho_compressed_2009}. This ensures the reconstruction is not solely dependent on the inherent bias of \ac{CNN} to produce noise-free images. Rather, well-known image characteristics that are embedded into \ac{CS} algorithms are also leveraged. We emphasize that the proposed technique is compatible with existing denoising \ac{CNN} and \ac{CS}-\ac{MRI} algorithms by deploying the popular \ac{DcCNN}~\cite{schlemper_deep_2017} as the \ac{CNN} to be trained on a single image, and \ac{D-AMP}~\cite{metzler_denoising_2016} as the \ac{CS} technique. 

\section{Methods}
\label{deepmri:methods}

\ac{CS}-\ac{MRI} is often an iterative solution to,
\begin{equation}
    \min_\mathbf{x} \| F_\Omega\mathbf{x} - \mathbf{y}_\Omega \|_2^2 + \lambda \mathcal{R}(\mathbf{x}), \label{eq:cs}
\end{equation}
where $F_\Omega$ represents the masked \ac{2D} \ac{DFT} at locations $\Omega$, $\mathbf{y}_\Omega$ is under-sampled $k$-space and $\mathcal{R}(\cdot)$ is some regularisation function~\cite{lustig_sparse_2007}. This work will investigate \ac{D-AMP} and \ac{DcCNN} for use in our \ac{SS}-\ac{CS}-\ac{MRI}. 

\ac{D-AMP} is a \ac{CS} algorithm that replaces $\mathcal{R}(\cdot)$ in Eq.~\ref{eq:cs} with the prior inherent to Gaussian denoisers~\cite{metzler_denoising_2016}. Construction of \ac{D-AMP} begins by considering the \ac{ISTA} solution to Eq.~\ref{eq:cs},
\begin{equation}
    \begin{split}
    \mathbf{r}_t &= \hat{\mathbf{x}}_{t} + F_\Omega^H(\mathbf{y}_\Omega - F_\Omega \hat{\mathbf{x}}_{t}), \\
    \hat{\mathbf{x}}_{t + 1} &= \argmin_\mathbf{x} \frac{1}{2} \|\mathbf{x} - \mathbf{r}_{t}\|_2^2 + \lambda_t \mathcal{R}(\mathbf{x}).
    \end{split}
    \label{eq:deep-istaupdate}
\end{equation}
It is easy to see that the $\argmin_\mathbf{x} \frac{1}{2} \|\mathbf{x} - \mathbf{r}_t\|_2^2 + \lambda_t\mathcal{R}(\mathbf{x})$ term resembles the \ac{MAP} estimation of $\mathbf{r}$. If we assume the residual $\mathbf{x} - \mathbf{r}_t$ has Gaussian distribution, then one could simply deploy existing Gaussian image denoising algorithms such as \ac{BM3D} to regularise the solution,
\begin{equation}
    \hat{\mathbf{x}}_{t + 1} = D_\sigma(\mathbf{r}_t),
\end{equation}
with $D_\sigma(\cdot)$ being the chosen Gaussian denoiser and $\sigma$ a hyper-parameter for standard deviation of noise. Unfortunately, successive calls to $D_\sigma(\cdot)$ biases intermediate solutions to remove ```observable'' Gaussian noise in the residual, limiting the effectiveness of Gaussian denoising and slowing convergence~\cite{metzler_denoising_2016}. \Ac{D-AMP} not only ensures the residual retains a Gaussian distribution between iterations, it is also capable of estimating the standard deviation of noise $\hat{\sigma}_t$ by means of the Onsager correction term~\cite{metzler_denoising_2016}. In the context of \ac{MRI} and Eq.~\ref{eq:deep-istaupdate}, then $F_\Omega (\mathbf{r}_{t-1} - \hat{\mathbf{x}}_{t-1}) \text{div} D_{\hat{\sigma}_{t-1}}(\mathbf{r}_{t-1}) / m$ is the Onsager correction term and $m$ is the total number of sampled points in $\Omega$. We have implemented this algorithm using Python's NumPy and the BM3D denoising algorithm.

\ac{DcCNN} follows a \ac{DL}-\ac{CS} paradigm that unrolls an iterative reconstruction scheme into a series of \ac{CNN} denoising and \ac{DC} steps. We implement the network as follows,
\begin{equation}
    \begin{split}
        \mathbf{x}_{t} &= f(\hat{\mathbf{x}}_{t} | \theta_t) \\
        \hat{\mathbf{x}}_{t + 1} &= F^H(F_{\bcancel{\Omega}}\mathbf{x}_t \cup \mathbf{y}_\Omega)
    \end{split} \label{eq:dccnn}
\end{equation}
where $f(\cdot|\theta_t)$ is the denoising \ac{CNN} at iteration $t$ with weights $\theta_t$, $\bcancel{\Omega}$ indicates the set of points not in $\Omega$. \ac{DC} therefore replaces $k$-space coefficients of $\mathbf{x}_{t}$ with known measurements $\mathbf{y}_\Omega$ in each iteration, whilst keeping the \ac{CNN} generated samples in $\bcancel{\Omega}$. We express a call to this network as,
\begin{equation}
    \hat{\mathbf{x}}_\Omega = f(F_\Omega^H\mathbf{y}_\Omega | \Theta),
\end{equation}
where $\Theta = \{\theta_t\}_{t=1}^n$ is the set of network weights in all iterations and $F_\Omega^H \mathbf{y}_\Omega$ is the \ac{ZF} estimate. The authors originally propose to train this network by computing the \ac{MSE} between reconstructed and ground truth images within a large dataset. We instead propose a deep and sparse single image \ac{SS}-\ac{CS}-\ac{MRI} loss (see Fig.~\ref{fig:self-supervise}).

% ConvDecoder~\cite{darestani_accelerated_2021} is a DIP-style approach to \ac{CS}-\ac{MRI}, where a random latent vector $\mathbf{z}$ is mapped to a single target \ac{MR} image. Inference is performed via, 
% \begin{equation}
%     \hat{\mathbf{x}} = f(\mathbf{z} | \Theta),
% \end{equation}
% with loss function,
% \begin{equation}
%     \min_\Theta \; \|\mathbf{y}_\Omega - F_{\Omega}f(\mathbf{z} | \Theta)\|_2^2. \label{eq:convdec}
% \end{equation}
% This procedure is performed for several epochs, where the model with the lowest loss is used for the final reconstruction, followed by a single \ac{DC} operation. A \ac{DIP} approach such as this has two significant drawbacks compared to other \ac{SS} approaches.
% \begin{enumerate}
%     \item Eq.~\ref{eq:convdec} simply maps the output image to known measurements, therefore, a valid solution could be the zero-filled and corrupted under-sampled image.
%     \item ConvDecoder relies entirely on the inductive bias of \ac{CNN} to produce noise-free images for a good reconstruction.
% \end{enumerate}
% To address these concerns, this chapter proposes a DeepRED-style reconstruction approach that unrolls an iterative recovery scheme into a \ac{CNN}, along with a robust \ac{SS} procedure. The reconstruction is additionally supervised by a hand-crafted \ac{CS}-\ac{MRI} algorithm. This not only leverages the inductive biases of each \ac{CNN} layer to remove \ac{CS} artefacts, but further incorporates prior knowledge about the target image. 

\subsection{Self-Supervised DL-MRI}
\label{deepmri:selfsup}

Similar to \ac{ZS-SSL}~\cite{yaman_zero-shot_2022}, our \ac{SS} uses the corrupted image and knowledge of the degradation process (under-sampled $k$-space) to train an image recovery model. Each epoch, $\Omega$ is divided into randomly generated $\Lambda$ and $\Upsilon$, such that $\Omega = \Lambda \cup \Upsilon$; \ac{ZS-SSL} uses a fixed $K$ pairs of $\Lambda$ and $\Upsilon$ with an additional `validation' set $\Gamma$ to detect over-fitting. The \ac{CNN} denoiser $f(\cdot | \Theta)$ is therefore trained to recover $\Upsilon$ from $\Lambda$, ensuring network outputs from each randomly generated subset are projected onto the same image. Due to the bias of \ac{CNN} to produce noise-free images~\cite{ulyanov_deep_2018}, it is hoped that the reconstruction converges to the original fully-sampled image rather than the \ac{ZF} solution. The ``deep'' component of our \ac{SS} loss is therefore defined as, 
\begin{equation}
    L_{kdc} := \| \mathbf{y}_\Upsilon - F_\Upsilon \mathbf{\hat{x}}_\Lambda \|_2^2 \label{eq:kdc}
\end{equation}
where,
\begin{equation}
    \mathbf{\hat{x}}_\Lambda = f(F_\Lambda^H\mathbf{y}_\Lambda | \Theta).  \label{eq:flamb}
\end{equation}
While Eq.~\ref{eq:kdc} is still prone to over-fitting to the \ac{ZF} data, \ac{ZS-SSL}~\cite{yaman_zero-shot_2022} demonstrated that this single-image \ac{SS} paradigm results in higher-quality images than the original \ac{DIP}~\cite{ulyanov_deep_2018} (and ConvDecoder~\cite{darestani_accelerated_2021}). We also find that randomly dividing $\Omega$ each epoch rather than using a fixed set of $K$ training pairs improved convergence characteristics and removed the need for an additional validation set.

% Here we will discus the mechanisms by which updates can be made via deep red
\subsection{Self-Supervised DL-CS-MRI}
\label{deepmri:deepmri}

To leverage the non-linearity of \ac{CNN} for single image applications and ensure the model does not converge to the \ac{ZF} solution, we improve \ac{SS} via the following DeepRED-style optimisation function,
\begin{align}
    \min_{\mathbf{x}, \Theta} \frac{1}{2} L_{kdc} + \frac{\lambda}{2}\mathbf{x}^H(\mathbf{x} - g(\mathbf{x}, \mathbf{y}_\Omega)) \; \; \; s.t. \; \mathbf{x} = f(\cdot|\Theta) \label{eq:deepred}
\end{align}
where $g(\cdot)$ is a denoiser of choice; other variables are defined in Section~\ref{deepmri:selfsup}. The \ac{SS} loss $L_{kdc}$ (Eq.~\ref{eq:kdc}) ensures fidelity between outputs of $f(\cdot | \Theta)$ and known measurements $\mathbf{y}_\Omega$, whereas $\frac{\lambda}{2}\mathbf{x}^H(\mathbf{x} - g(\mathbf{x}, \mathbf{y}_\Omega))$ constrains the output image to solutions for which $g(\cdot)$ has little impact; if $g(\cdot)$ is a Gaussian image denoiser the loss encourages $\mathbf{x}$ to be noise-free. Similar to DeepRED~\cite{mataev_deepred_2019}, we propose an \ac{ADMM} approach, where Eq.~\ref{eq:deepred} is split into two sub-problems. \newline

\noindent
\textbf{Step 1: Update CNN Weights}
By fixing $\mathbf{x}_k$ and Lagrange multipliers vector $\mathbf{q}_k$ at iteration $k$, the \ac{CNN} training loss becomes,
\begin{equation}
    \Theta_{k + 1} = \arg\min_\Theta \frac{1}{2} L_{kdc} + \frac{\mu}{2} \|\mathbf{x}_k - \hat{\mathbf{x}}_\Lambda - \mathbf{q}_k \|_2^2. \label{eq:cnn}
\end{equation}
Here, $\mathbf{x}_k$ is the current denoiser image estimate and $\hat{\mathbf{x}}_\Lambda$ is the \ac{CNN} estimate as-per Eq.~\ref{eq:flamb}. Inclusion of the $L_{CS} := \frac{\mu}{2} \|\mathbf{x}_k - \hat{\mathbf{x}}_\Lambda - \mathbf{q}_k \|_2^2$ term improves upon \ac{SS} by incorporating information from the chosen denoiser function ($L_{CS}$ in Fig.~\ref{fig:self-supervise}). As Eq.~\ref{eq:cnn} resembles the objective function of many \ac{DNN} problems, updates to $\Theta$ can be computed via back-propagation. \newline

\noindent
\textbf{Step 2: Denoise Estimate}
Fixing $\Theta_k$, the \ac{RED} objective is,
\begin{align}
    \mathbf{x}_{k+1} = \arg\min_\mathbf{x} \frac{\lambda}{2} \mathbf{x}^H(\mathbf{x} - g(\mathbf{x}, \mathbf{y}_\Omega)) + \frac{\mu}{2} \|\mathbf{x} - \hat{\mathbf{x}}_\Omega - \mathbf{q}_k \|_2^2 \label{eq:red}
\end{align}
which is solved via the fixed-point strategy, where the derivative of Eq.~\ref{eq:red} is zeroed with respect to $\mathbf{x}$ and leads to,
\begin{equation}
    \mathbf{x}_{k+1} = \frac{1}{\lambda + \mu} (\lambda g(\hat{\mathbf{x}}_\Omega, \mathbf{y}_\Omega) + \mu (\hat{\mathbf{x}}_\Omega + \mathbf{q}_k))
\end{equation}
Lagrange multipliers $\mathbf{q}$ is updated via, $\mathbf{q}_{k + 1} = \mathbf{q}_k + \eta(\hat{\mathbf{x}}_\Omega - \mathbf{x}_{k})$. While simply setting the denoising function $g(\cdot)$ to a Gaussian denoiser may seem suitable at first glance, the non-local and structured nature of \ac{CS} artefacts pose problems for Gaussian models. Additionally, choice of denoiser standard deviation of noise ($\sigma$) is an important hyperparameter to set. For this reason, we have implemented \ac{D-AMP} as described in Section~\ref{deepmri:methods}, which leverages the Onsager correction term to estimate the remaining standard deviation of noise ($\hat{\sigma}$) in a given iteration. While this approach requires several computations of the denoiser internal to \ac{D-AMP}, it removes a hyperparameter from the overall algorithm and can be computed in parallel to the \ac{CNN} training step. Minimising Eq.~\ref{eq:deepred} in this two-step manner will produce an output image $\mathbf{x}$ that is consistent with both the degraded measurements $\mathbf{y}_\Omega$ and the denoised image from $g(\mathbf{x}, \mathbf{y}_\Omega)$.

\subsection{Implementation Details}
\label{deepmri:implementation}

% \Ac{SS} is \ac{DcCNN} trained only with $L_{kdc}$ (Eq.~\ref{eq:kdc}) and represents \ac{ZS-SSL} without additional validation subset. \Ac{SS}-\ac{D-AMP} implements the proposed deep \ac{CS} methodology (Eq.~\ref{eq:deepred}), where the image estimate $\hat{\mathbf{x}}_\Omega$ at epoch $1100$ is fed into \ac{D-AMP} and returned at epoch $2100$ . \Ac{SS}-\ac{BM3D} replaces $g(\cdot)$ with \ac{BM3D} and a fixed $\sigma=0.012$ applied every 30 epochs. Metrics are \ac{PSNR} and \ac{SSIM}.

7 central slices from the 10 test Calgary-Campinas T1 brain~\cite{souza_open_2018} ($256 \times 256$) volumes and 10 randomly chosen validation volumes from the NYU fastMRI PD knee~\cite{knoll_fastmri_2020} ($320 \times 320$) datasets have been used for testing, providing complex-valued 3T \ac{2D} $k$-space from an emulated single-coil methodology. This gives 70 slices for each anatomy, allowing for computation in reasonable time. 

\begin{figure}[t!]
    \small
    \begin{minipage}[t]{\linewidth}
        \centering
        \captionof{table}{Parameters used in experiments (found via grid-search). In SS-BM3D, denoiser standard deviation of noise is normalised between $\sigma \in [0, 1]$. We deploy $\sigma=0.012$. ConvDecoder parameters are as-per the original implementation~\cite{darestani_accelerated_2021}.}
        \label{tab:params-deep}
        \begin{tabularx}{\linewidth}{c||Y|Y|Y|Y|Y|Y|Y|Y}
            CS        & $\lambda$ & $\mu$ & $\eta$ & CS iter. & CS int. & CNN iter. & SS iter. & LR ($10^{-3}$)               \\ \hline \hline
             \multicolumn{9}{c}{\textbf{Brains}}   \\ \hline
            ConvDec~\cite{darestani_accelerated_2021} & -         & - & -    & -        & -       & -           & 5000   & 8   \\ \hline
            SS        & -         & -     & -       & -         & -       & 7           & 4000     & 1     \\ \hline
            SS-BM3D   & 0.125       & 0.25   & 0.001       & 1       & 30      & 7           & 4000     & 1       \\ \hline
            SS-D-AMP  & 3         & 1     & 0.001       & 25         & 1000     & 7           & 4000     & 1     \\ \hline\hline
             \multicolumn{9}{c}{\textbf{Knees}}   \\ \hline
            ConvDec~\cite{darestani_accelerated_2021} & -         & - & -    & -        & -       & -           & 5000   & 8   \\ \hline
            SS        & -         & -     & -       & -         & -       & 7           & 3000     & 1     \\ \hline
            SS-BM3D   & 0.5       & 1.0   & 1       & 1       & 30      & 7           & 3000     & 1       \\ \hline
            SS-D-AMP  & 1.5         & 0.5     & 0.001       & 25         & 1000     & 7           & 3000     & 1     \\\hline
        \end{tabularx}
    \end{minipage}
    \par\medskip
    \begin{minipage}[b]{\linewidth}
        \centering
        \begin{subfigure}[t]{\linewidth}
            \centering
            \includegraphics[width=0.24\linewidth, angle=270]{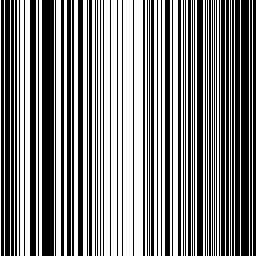}
            \hfill
            \includegraphics[width=0.24\linewidth, angle=270]{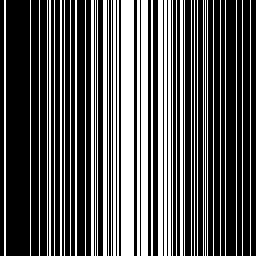}
            \hfill
            \includegraphics[width=0.24\linewidth, angle=270]{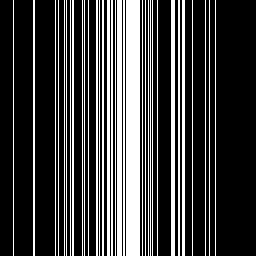}
            \hfill
            \includegraphics[width=0.24\linewidth, angle=270]{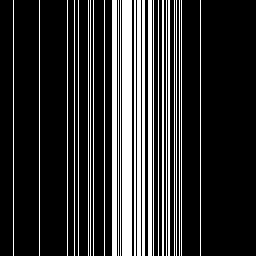}
        \end{subfigure}
        \par\smallskip
        \begin{subfigure}[b]{\linewidth}
            \centering
            \includegraphics[width=0.24\linewidth]{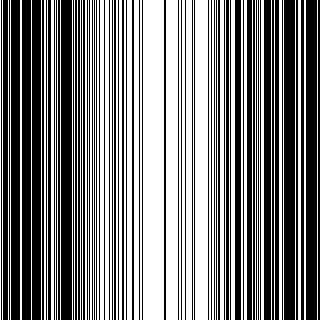}
            \hfill
            \includegraphics[width=0.24\linewidth]{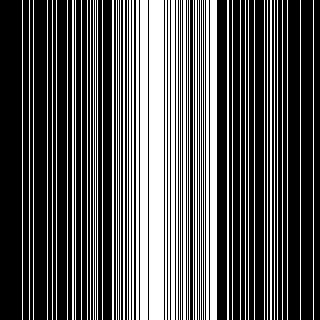}
            \hfill
            \includegraphics[width=0.24\linewidth]{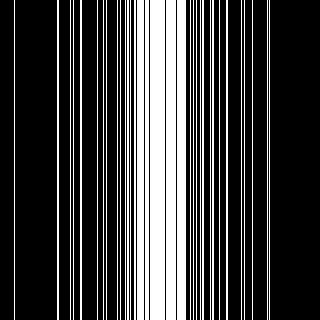}
            \hfill
            \includegraphics[width=0.24\linewidth]{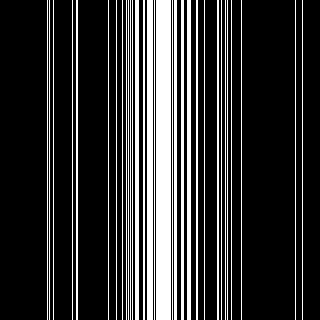}
        \end{subfigure}
        \caption{Sampling masks used in this study. Brains (top) and knees (bottom). Left to right: R2, R3, R4, R5.}
    \end{minipage}
\end{figure}

Training step (Eq.~\ref{eq:cnn}) is performed on an NVIDIA SMX-2T V100. All other related computation (e.g. BM3D, D-AMP) is performed on 28 cores (56 threads) of an Intel Xeon 6132. \ac{DcCNN} was implemented using PyTorch and trained using the Adam optimiser. Subsets $\Lambda$ and $\Upsilon$ are randomly generated each epoch, each containing $50\%$ of available samples and a fully-sampled $20\times 20$ region of central $k$-space. Table~\ref{tab:params-deep} summarises the chosen hyperparameters. SS indicates \ac{DcCNN} trained only on the \ac{SS} loss ($L_{kdc}$ in Eq.~\ref{eq:kdc}). This is similar to the training procedure proposed by \ac{ZS-SSL}, however, we find that continuously generating random samples of $\Lambda, \Upsilon$ helps to avoid over-fitting. As such, a separate validation subset was not necessary in our testing. SS-D-AMP implements the proposed \ac{SS}-\ac{CS}-\ac{MRI} methodology (Eq.~\ref{eq:deepred}). SS-BM3D replaces $g(\cdot)$ with just the BM3D denoiser and a fixed $\sigma=0.012$. \ac{CS} iter. is the number of \ac{CS} iterations in $g(\cdot)$, \ac{CS} int. is the number of SS iterations (epochs) between \ac{CS} updates, and \ac{DcCNN} iter. relates to the number of cascaded \ac{CNN} blocks. It should be noted, \ac{SS}-\ac{D-AMP} operates exactly as \ac{SS} for the first $2100$ epochs, in that only Eq.~\ref{eq:kdc} is used for training. At epoch $1100$, the \ac{CNN} estimate ($\hat{\mathbf{x}}_\Omega = f(F_\Omega^H\mathbf{y}_\Omega | \Theta_{1100})$) is fed into \ac{D-AMP} $g(\hat{\mathbf{x}}_\Omega, \mathbf{y}_\Omega)$. This is evaluated in parallel for $1000$ epochs, where the result can be used to update $\mathbf{x}$ and the network begins training using Eq.~\ref{eq:cnn}. The first instance of this is visualised by the red arrows in Fig.~\ref{fig:plot}. Sampling masks were randomly generated using an $\ell_1$ $k$-space distribution, then fixed for each anatomy and reduction factor. The metrics used are \ac{PSNR} and \ac{SSIM}. 

\section{Results and Discussion}

\begin{figure}[t!]
    \centering
    \includegraphics[height=0.48\linewidth, angle=90]{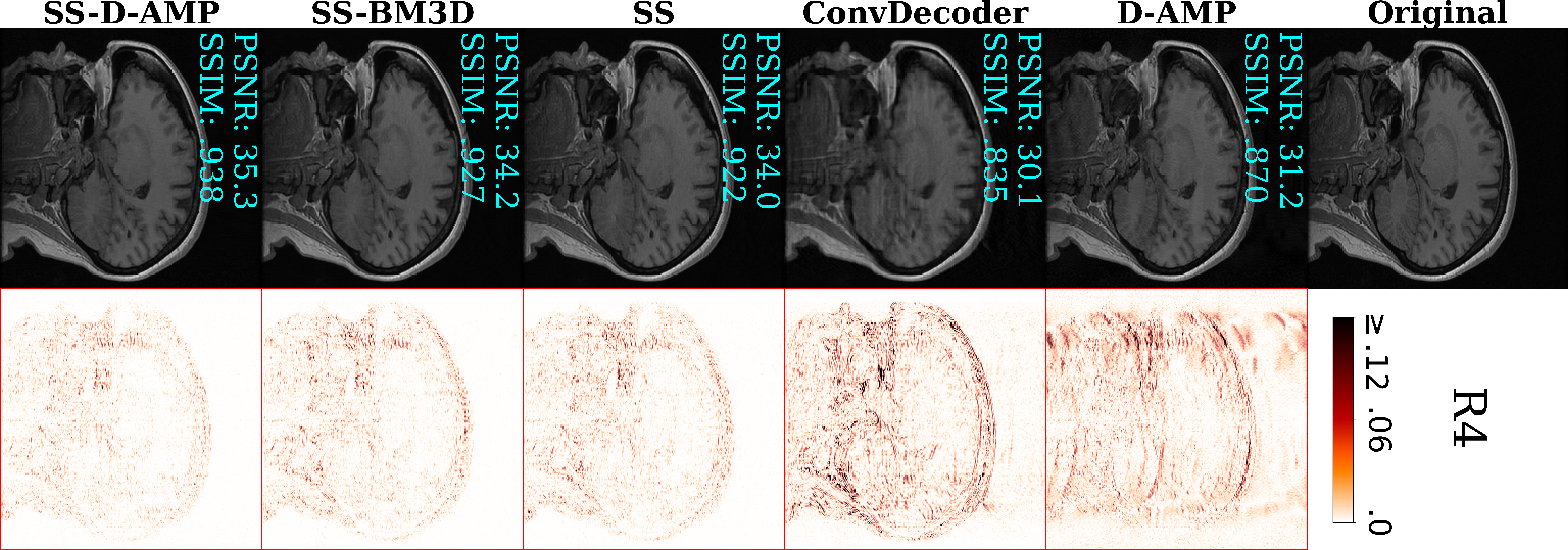}
    \includegraphics[height=0.4555\linewidth, angle=90]{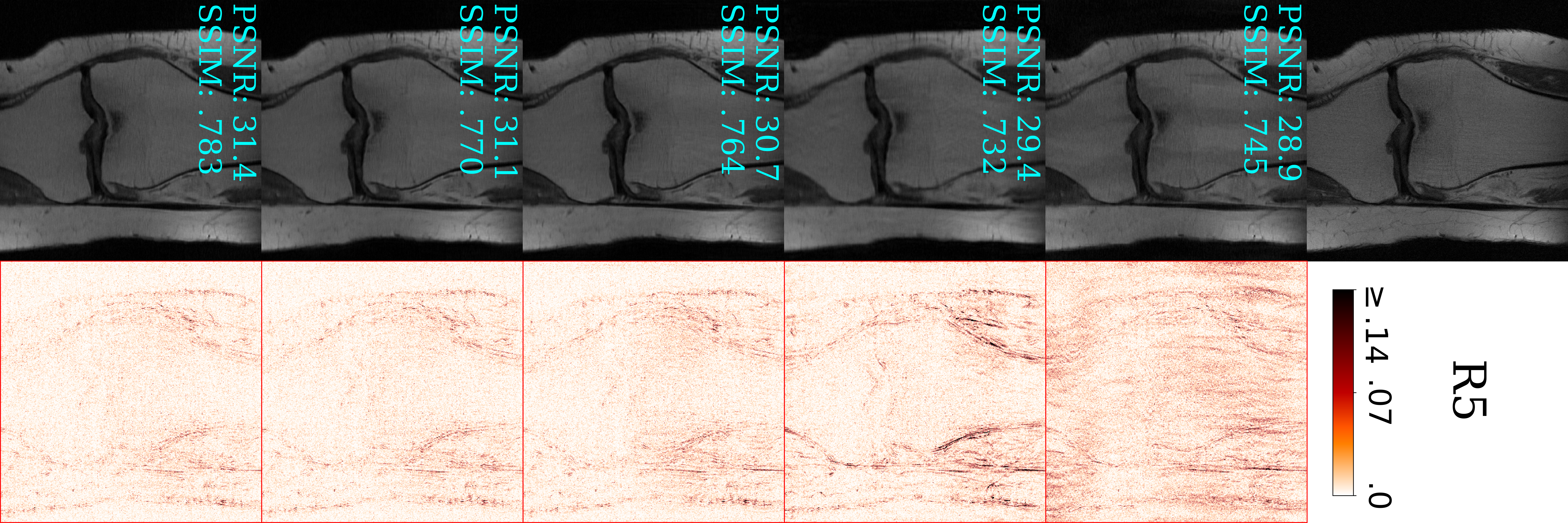}
    \caption{Representative reconstructions, comparing performance from our \ac{SS}-based methods on the brain~\cite{souza_open_2018} and knee~\cite{knoll_fastmri_2020} datasets. \ac{SS} alone is indicative of \ac{ZS-SSL}~\cite{yaman_zero-shot_2022}.}
    \label{fig:qualitative}
\end{figure}

\noindent
\textbf{Qualitative Results:} % automatically adjusts the denoising to avoid over-smoothing
Fig.~\ref{fig:qualitative} contains a representative comparison between existing single-image techniques \ac{D-AMP}~\cite{metzler_denoising_2016} and ConvDecoder~\cite{darestani_accelerated_2021} to our proposed approach (best viewed zoomed in). \ac{D-AMP} does well to remove localised artefacts, however, it struggles with non-local and structured perturbations. ConvDecoder, and to a lesser extent \ac{SS}, produce localised ``ringing'' artefacts in the direction of under-sampling. Compared to \ac{D-AMP} and ConvDecoder, \ac{SS} effectively penalises non-local and structured \ac{CS} artefacts, with \ac{SS}-\ac{BM3D} scoring higher in terms of \ac{PSNR} and \ac{SSIM} (reflected in the error map). Though, \ac{SS}-\ac{BM3D} does not meaningfully dampen ringing artefacts as they are not necessarily noise-like. We find \ac{SS}-\ac{D-AMP} features the sharpest images with less visible artefacts and better preserved textures, effectively combining the strengths of \ac{D-AMP} and \ac{SS}. \newline

\begin{figure}[t]
    \begin{minipage}[t]{\linewidth}
        \centering
        % \footnotesize
        \small
        \captionof{table}{Average performance achieved by each reconstruction method per dataset. PSNR (dB) SSIM (\%).}
        \label{tab:results}
        \begin{tabularx}{\linewidth}{c|YY|YY|YY|YY|c}
        \hline
        \multirow{2}{*}{\textbf{Method}} & \multicolumn{2}{c|}{\textbf{R=2}} & \multicolumn{2}{c|}{\textbf{R=3}} & \multicolumn{2}{c|}{\textbf{R=4}} & \multicolumn{2}{c|}{\textbf{R=5}} & \textbf{s/}  \\ \cline{2-9}
                                         & PSNR            & SSIM            & PSNR            & SSIM            & PSNR            & SSIM            & PSNR            & SSIM            & \textbf{img} \\ \hline \hline
        \multicolumn{10}{c}{\textbf{Brains}}                                                                                                                                                            \\ \hline
        Zero-Fill                        & 28.10           & 0.784           & 25.89           & 0.720           & 25.34           & 0.689           & 24.81           & 0.672           & -            \\
        D-AMP                            & 38.74           & 0.971           & 32.29           & 0.904           & 31.06           & 0.874           & 30.50           & 0.861           & \textbf{68}  \\
        ConvDec                          & 35.93           & 0.941           & 32.14           & 0.888           & 30.28           & 0.845           & 29.58           & 0.836           & {\ul 130}    \\
        SS                               & 42.56           & 0.983           & 37.08           & 0.957           & {\ul 34.17}     & 0.927           & {\ul 32.45}     & 0.907           & 209          \\ \hline
        SS-BM3D                          & {\ul 42.68}     & {\ul 0.983}     & {\ul 37.15}     & {\ul 0.958}     & 34.15           & {\ul 0.931}     & 32.34           & {\ul 0.912}     & 329          \\
        SS-D-AMP                         & \textbf{43.72}  & \textbf{0.986}  & \textbf{38.25}  & \textbf{0.965}  & \textbf{35.24}  & \textbf{0.941}  & \textbf{33.34}  & \textbf{0.921}  & 275          \\ \hline \hline
        \multicolumn{10}{c}{\textbf{Knees}}                                                                                                                                                             \\ \hline
        Zero-Fill                        & 27.74           & 0.778           & 27.00           & 0.719           & 27.87           & 0.706           & 26.34           & 0.658           & -            \\
        D-AMP                            & 30.53           & 0.849           & 30.07           & 0.804           & 30.60           & 0.781           & 29.29           & 0.746           & \textbf{78}  \\
        ConvDec                          & 33.15           & 0.860           & 31.53           & 0.798           & 30.92           & 0.760           & 29.71           & 0.724           & {\ul 190}    \\
        SS                               & 34.51           & 0.880           & 32.74           & 0.821           & 31.74           & 0.777           & 30.93           & 0.748           & 232          \\ \hline
        SS-BM3D                          & {\ul 34.52}     & {\ul 0.881}     & {\ul 32.79}     & {\ul 0.822}     & {\ul 31.79}     & {\ul 0.778}     & {\ul 31.02}     & {\ul 0.750}     & 300          \\
        SS-D-AMP                         & \textbf{34.65}  & \textbf{0.884}  & \textbf{32.97}  & \textbf{0.829}  & \textbf{32.04}  & \textbf{0.792}  & \textbf{31.33}  & \textbf{0.767}  & 254          \\ \hline
        \end{tabularx}
    \end{minipage}
    \begin{minipage}[b]{\linewidth}
        \centering
        \begin{subfigure}[l]{0.49\linewidth}
            \includegraphics[width=\linewidth]{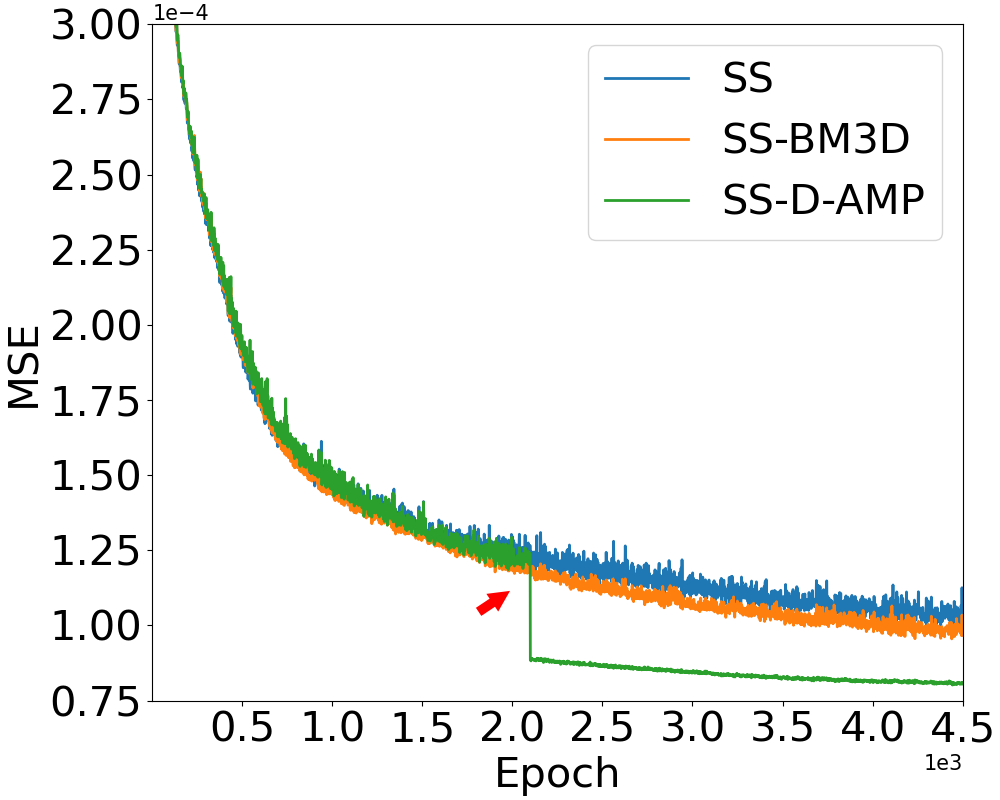}
        \end{subfigure}
        \begin{subfigure}[r]{0.49\linewidth}
            \includegraphics[width=\linewidth]{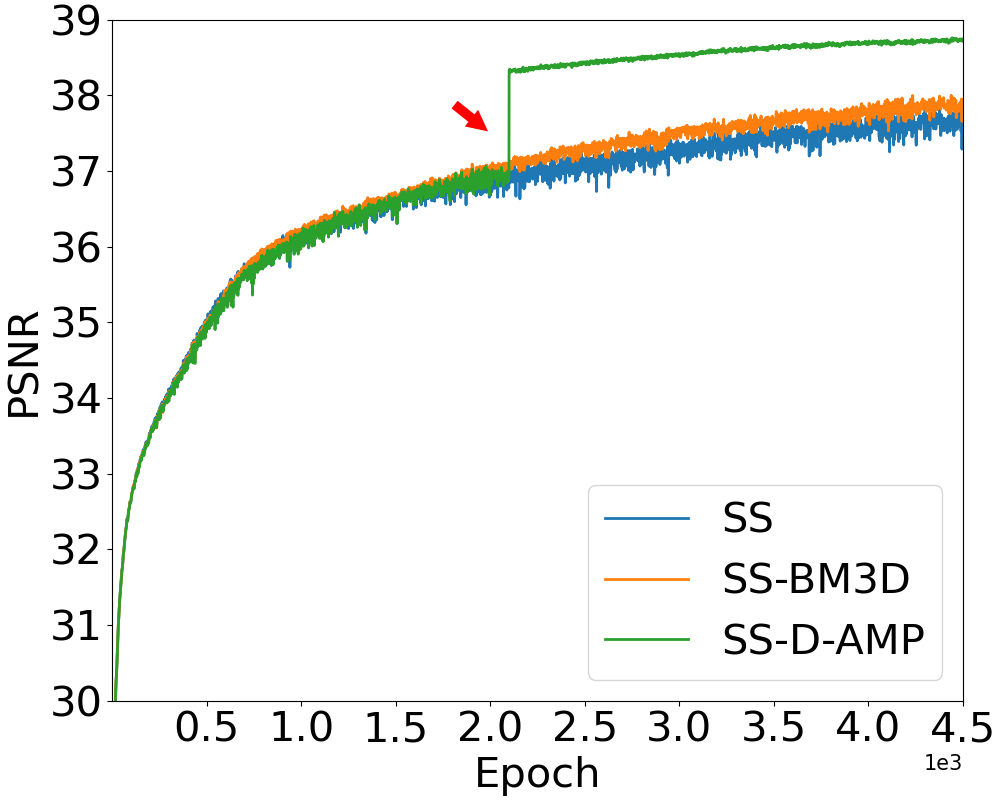}
        \end{subfigure}
        \captionof{figure}{Representative test loss. Continuously randomizing $\Lambda, \Upsilon$ helps avoid overfitting. \Ac{SS}-\ac{D-AMP} improves upon \ac{SS} by stabilising epoch-to-epoch variations and incorporating the \ac{D-AMP} reconstruction into the loss function (shown by red arrows).}
        \label{fig:plot}
    \end{minipage}
\end{figure}

\noindent
\textbf{Quantitative Results:}
Fig.~\ref{fig:plot} illustrates the test loss convergence characteristics of a representative brain reconstruction. Blue is \ac{SS}, where only Eq.~\ref{eq:kdc} is used for training (similar to \ac{ZS-SSL}~\cite{yaman_zero-shot_2022}). Orange is \ac{SS}-\ac{BM3D} and green is \ac{SS}-\ac{D-AMP}, each replacing $g(\cdot)$ with \ac{BM3D} and \ac{D-AMP} respectively. While \ac{SS}-\ac{BM3D}'s constant denoising (every $30$ epochs) gradually improves the reconstruction compared to \ac{SS}, denoiser estimates for \ac{SS}-\ac{D-AMP} are only completed every $1000$ epochs. In the location of the red arrows, it can be seen that \ac{SS}-\ac{D-AMP} noticeably reduces variation between epochs and allows the overall image estimate to improve by $1.5$dB compared to \ac{SS}. This is because \ac{D-AMP} regularises $\hat{\mathbf{x}}_\Omega$ in a manner that ensures \ac{CS} artefacts are removed without over-smoothing~\cite{metzler_denoising_2016}. As a result, $f(\cdot|\Theta)$ continues to minimize the $k$-space consistency loss, whilst ensuring $\mathbf{x}$ does not diverge from the \ac{CS} estimate.

Table~\ref{tab:results} contains the average performance achieved on the 70 brain and 70 knee test slices. \ac{SS}-\ac{D-AMP} scores highest overall, improving upon \ac{D-AMP} by an average 4.49dB and 2.63dB for brains and knees respectively. The main penalty to \ac{SS}-\ac{D-AMP} is the increased computation compared to \ac{D-AMP} and \ac{SS}. However, Fig.~\ref{fig:plot} indicates that the \ac{SS}-\ac{D-AMP} estimate at epoch $2100$ would produce an image with superior image quality compared to \ac{SS} or \ac{SS}-\ac{BM3D}, enabling a reduction in run-time. Further, \ac{SS}-\ac{BM3D}'s non-adaptive Gaussian denoising is less efficient at dampening \ac{CS}-\ac{MRI} artefacts than \ac{SS}-\ac{D-AMP}, requiring more computation time overall. As ConvDecoder is limited to the mapping of a random vector $\mathbf{z}$ to known measurements $\mathbf{y}_\Omega$, the reconstruction sees some overfitting and scores lowest on the brains.

\section{Discussion and Conclusion}

Hand-crafted \ac{CS}-\ac{MRI} algorithms provide reliable and robust solutions to the image recovery problem. However, \ac{DL} has emerged as a promising direction due to its ability to produce state-of-the-art images. Unfortunately, \ac{DL} typically requires large datasets to train satisfactory solutions. Additionally, generalisability across multiple datasets is not guaranteed. Alternatively, single image \ac{SS}-\ac{DL} techniques such as \ac{ZS-SSL} are capable of high-quality image recovery by leveraging the relationship between captured data and the image degradation model. However, reliance on the inherent bias of \ac{CNN} to produce noise-free images is prone to over-fitting~\cite{yaman_zero-shot_2022}. Recognising that desirable image characteristics are embedded into hand-crafted algorithms, this work develops a \ac{SS}-\ac{CS}-\ac{MRI} paradigm that successfully combines \ac{CS} algorithms with \ac{DL} to ensure the solution adheres to well-established image requirements (such as image sparsity). Our findings suggest that this deep and sparse reconstruction is well suited to \ac{CS}-\ac{MRI}, outperforming traditional hand-crafted and single image \ac{SS}-\ac{DL} techniques. Future work should consider integrating other domain knowledge into the reconstruction procedure, such as leveraging redundancies in multi-coil data (i.e. low-rank). Additionally, inference time can be improved via GPU implementations of \ac{D-AMP} and \ac{BM3D}; due to the \ac{PnP} nature of our approach, $g(\cdot)$ can also be replaced by pre-trained \ac{DL}-\ac{CS} or Gaussian denoisers. As demonstrated in~\cite{darestani_accelerated_2021, yaman_zero-shot_2022}, pre-trained weights $\Theta$ could also be fine-tuned on a single image using our \ac{SS}-\ac{CS}-\ac{MRI}. 

\section{Compliance with Ethical Standards}
This research study was conducted retrospectively using human subject data made available in open access by~\cite{souza_open_2018, knoll_fastmri_2020}. Ethical approval was not required as confirmed by the license attached with the open access data.

% \small
\bibliographystyle{IEEEbib}
\sloppy
\bibliography{MICCAI}

\begin{thebibliography}{10}

\bibitem{lustig_sparse_2007}
Michael Lustig, David Donoho, and John~M. Pauly,
\newblock ``Sparse {MRI}: {The} application of compressed sensing for rapid
  {MR} imaging,''
\newblock {\em Magnetic Resonance in Medicine}, vol. 58, no. 6, pp. 1182--1195,
  2007.

\bibitem{metzler_denoising_2016}
Christopher~A. Metzler, Arian Maleki, and Richard~G. Baraniuk,
\newblock ``From {Denoising} to {Compressed} {Sensing},''
\newblock {\em IEEE Transactions on Information Theory}, vol. 62, no. 9, pp.
  5117--5144, Sept. 2016.

\bibitem{ye_compressed_2019}
Jong~Chul Ye,
\newblock ``Compressed sensing {MRI}: a review from signal processing
  perspective,''
\newblock {\em BMC Biomedical Engineering}, vol. 1, no. 1, pp. 8, Dec. 2019.

\bibitem{schlemper_deep_2017}
Jo~Schlemper, Jose Caballero, Joseph~V. Hajnal, Anthony Price, and Daniel
  Rueckert,
\newblock ``A {Deep} {Cascade} of {Convolutional} {Neural} {Networks} for {MR}
  {Image} {Reconstruction},''
\newblock in {\em Information {Processing} in {Medical} {Imaging}}, Marc
  Niethammer, Martin Styner, Stephen Aylward, Hongtu Zhu, Ipek Oguz, Pew-Thian
  Yap, and Dinggang Shen, Eds., Cham, 2017, Lecture {Notes} in {Computer}
  {Science}, pp. 647--658, Springer International Publishing.

\bibitem{chandra_deep_2021}
Shekhar~S Chandra, Marlon Bran~Lorenzana, Xinwen Liu, Siyu Liu, Steffen
  Bollmann, and Stuart Crozier,
\newblock ``Deep learning in magnetic resonance image reconstruction,''
\newblock {\em Journal of Medical Imaging and Radiation Oncology}, vol. 65, no.
  5, pp. 564--577, 2021.

\bibitem{lorenzana_aliasnet_2023}
Marlon {Bran Lorenzana}, Shekhar~S. Chandra, and Feng Liu,
\newblock ``{AliasNet}: Alias artefact suppression network for accelerated
  phase-encode {MRI},''
\newblock {\em Magnetic Resonance Imaging}, vol. 105, pp. 17--28, 2024.

\bibitem{yaman_self-supervised_2020}
Burhaneddin Yaman, Seyed Amir~Hossein Hosseini, Steen Moeller, Jutta Ellermann,
  Kâmil Uğurbil, and Mehmet Akçakaya,
\newblock ``Self-supervised learning of physics-guided reconstruction neural
  networks without fully sampled reference data,''
\newblock {\em Magnetic Resonance in Medicine}, vol. 84, no. 6, pp. 3172--3191,
  2020.

\bibitem{zeng_review_2021}
Gushan Zeng, Yi~Guo, Jiaying Zhan, Zi~Wang, Zongying Lai, Xiaofeng Du, Xiaobo
  Qu, and Di~Guo,
\newblock ``A review on deep learning {MRI} reconstruction without fully
  sampled k-space,''
\newblock {\em BMC Medical Imaging}, vol. 21, no. 1, pp. 195, Dec. 2021.

\bibitem{hu_self-supervised_2021}
Chen Hu, Cheng Li, Haifeng Wang, Qiegen Liu, Hairong Zheng, and Shanshan Wang,
\newblock ``Self-supervised {Learning} for {MRI} {Reconstruction} with a
  {Parallel} {Network} {Training} {Framework},''
\newblock in {\em Medical {Image} {Computing} and {Computer} {Assisted}
  {Intervention} – {MICCAI} 2021}, Marleen de~Bruijne, Philippe~C. Cattin,
  Stéphane Cotin, Nicolas Padoy, Stefanie Speidel, Yefeng Zheng, and Caroline
  Essert, Eds., Cham, 2021, Lecture {Notes} in {Computer} {Science}, pp.
  382--391, Springer International Publishing.

\bibitem{zhou_dual-domain_2022}
Bo~Zhou, Jo~Schlemper, Neel Dey, Seyed~Sadegh Mohseni~Salehi, Kevin Sheth, Chi
  Liu, James~S. Duncan, and Michal Sofka,
\newblock ``Dual-domain self-supervised learning for accelerated
  non-{Cartesian} {MRI} reconstruction,''
\newblock {\em Medical Image Analysis}, vol. 81, pp. 102538, Oct. 2022.

\bibitem{yaman_zero-shot_2022}
Burhaneddin Yaman, Seyed Amir~Hossein Hosseini, and Mehmet Akcakaya,
\newblock ``Zero-{Shot} {Self}-{Supervised} {Learning} for {MRI}
  {Reconstruction},''
\newblock in {\em International {Conference} on {Learning} {Representations}},
  2022.

\bibitem{ulyanov_deep_2018}
Dmitry Ulyanov, Andrea Vedaldi, and Victor Lempitsky,
\newblock ``Deep {Image} {Prior},''
\newblock 2018, pp. 9446--9454.

\bibitem{darestani_accelerated_2021}
Mohammad~Zalbagi Darestani and Reinhard Heckel,
\newblock ``Accelerated {MRI} {With} {Un}-{Trained} {Neural} {Networks},''
\newblock {\em IEEE Transactions on Computational Imaging}, vol. 7, pp.
  724--733, 2021.

\bibitem{quan_self2self_2020}
Yuhui Quan, Mingqin Chen, Tongyao Pang, and Hui Ji,
\newblock ``Self2self with dropout: Learning self-supervised denoising from
  single image,''
\newblock in {\em 2020 IEEE/CVF Conference on Computer Vision and Pattern
  Recognition (CVPR)}, 2020, pp. 1887--1895.

\bibitem{mataev_deepred_2019}
Gary Mataev, Peyman Milanfar, and Michael Elad,
\newblock ``{DeepRED}: {Deep} {Image} {Prior} {Powered} by {RED},''
\newblock 2019, pp. 0--0.

\bibitem{romano_little_2017}
Yaniv Romano, Michael Elad, and Peyman Milanfar,
\newblock ``The {Little} {Engine} {That} {Could}: {Regularization} by
  {Denoising} ({RED}),''
\newblock {\em SIAM Journal on Imaging Sciences}, vol. 10, no. 4, pp.
  1804--1844, 2017.

\bibitem{donoho_compressed_2009}
D.L. Donoho,
\newblock ``Compressed sensing,''
\newblock {\em IEEE Transactions on Information Theory}, vol. 52, no. 4, pp.
  1289--1306, 2006.

\bibitem{souza_open_2018}
Roberto Souza, Oeslle Lucena, Julia Garrafa, David Gobbi, Marina Saluzzi,
  Simone Appenzeller, Letícia Rittner, Richard Frayne, and Roberto Lotufo,
\newblock ``An open, multi-vendor, multi-field-strength brain {MR} dataset and
  analysis of publicly available skull stripping methods agreement,''
\newblock {\em NeuroImage}, vol. 170, pp. 482--494, Apr. 2018.

\bibitem{knoll_fastmri_2020}
Florian Knoll, Jure Zbontar, Anuroop Sriram, Matthew~J. Muckley, Mary Bruno,
  Aaron Defazio, Marc Parente, Krzysztof~J. Geras, Joe Katsnelson, Hersh
  Chandarana, Zizhao Zhang, Michal Drozdzalv, Adriana Romero, Michael Rabbat,
  Pascal Vincent, James Pinkerton, Duo Wang, Nafissa Yakubova, Erich Owens,
  C.~Lawrence Zitnick, Michael~P. Recht, Daniel~K. Sodickson, and Yvonne~W.
  Lui,
\newblock ``{fastMRI}: {A} {Publicly} {Available} {Raw} k-{Space} and {DICOM}
  {Dataset} of {Knee} {Images} for {Accelerated} {MR} {Image} {Reconstruction}
  {Using} {Machine} {Learning},''
\newblock {\em Radiology: Artificial Intelligence}, vol. 2, no. 1, pp. e190007,
  Jan. 2020.

\end{thebibliography}

\acrodef{CS}{compressed sensing}
\acrodef{MRI}{Magnetic resonance imaging}
\acrodef{MR}{magnetic resonance}
\acrodef{ADMM}{alternating directions method of multipliers}
\acrodef{D-AMP}{denoising-based approximate message passing}
\acrodef{ISTA}{iterative shrinkage-thresholding algorithm}
\acrodef{MAP}{maximum a posteriori}

\acrodef{ZF}{zero-filled}
\acrodef{ZS-SSL}{Zero-Shot Self-Supervised Learning}

\acrodef{DL}{deep learning}
\acrodef{CNN}{convolutional neural network}
\acrodef{DNN}{deep neural network}
\acrodef{TNN}{transformer neural network}
\acrodef{DcCNN}{Deep Cascade of Convolutional Neural Networks}
\acrodef{TL-MRI}{transform learning MRI}
\acrodef{DIP}{Deep Image Prior}
\acrodef{PnP}{plug-and-play}
\acrodef{RED}{regularisation by denoising}
\acrodef{PhaseNet}{phase reconstruction network}
\acrodef{MSE}{mean squared error}
\acrodef{MAE}{mean absolute error}

\acrodef{BM3D}{block-matching and 3D filtering}

\acrodef{1D}{one dimensional}
\acrodef{2D}{two dimensional}
\acrodef{DFT}{discrete Fourier transform}
\acrodef{DC}{data consistency}
\acrodef{PSNR}{peak signal to noise ratio}
\acrodef{SSIM}{structural similarity}
\acrodef{SS}{self-supervised}
\acrodef{PD}{proton density without fat suppression}
\end{document}